\documentclass[aps,twocolumn,10pt,pra,longbibliography]{revtex4-1}
\pdfoutput=1
\usepackage{amsmath}
\usepackage{latexsym}
\usepackage{graphicx}
\usepackage{color}
\usepackage{bm}
\usepackage{hyperref}
\usepackage[all]{hypcap}
\hypersetup{pdfauthor={Deny R. Hamel},pdftitle={Direct generation of three-photon polarization entanglement}}

\begin{document}

\title{Direct generation of three-photon polarization entanglement}
\author{Deny R. Hamel$^{1}$, Lynden K. Shalm$^{1,2}$, Hannes H\"{u}bel$^3$, Aaron J. Miller$^{2,4}$, Francesco  Marsili$^2$, Varun B. Verma$^2$, Richard P. Mirin$^2$,  Sae Woo Nam$^2$,  Kevin J. Resch$^1$ and Thomas Jennewein$^{1}$}
\affiliation{$^1$ Institute for Quantum Computing and Department of Physics \& Astronomy, \\ University of Waterloo,  Waterloo, N2L 3G1, Canada\\
         $^2$ National Institute of Standards and Technology, 325 Broadway St., Boulder CO 80305, USA\\
         $^3$ Physics Department, Stockholm University, S-10691 Stockholm, Sweden\\
         $^4$ Department of Physics, Albion College, Albion, MI 49224, USA.}

\begin{abstract}
Non-classical states of light are of fundamental importance for emerging quantum technologies.
All optics experiments producing multi-qubit entangled states have until now relied on outcome post-selection, a procedure where only the measurement results corresponding to the desired state are considered.
This method severely limits the usefulness of the resulting entangled states.
Here, we show the direct production of polarization-entangled photon triplets by cascading two entangled downconversion processes.
Detecting the triplets with high efficiency superconducting nanowire single-photon detectors allows us to fully characterize them through quantum state tomography.
We use our three-photon entangled state to demonstrate the ability to herald Bell states, a task which was not possible with previous three-photon states, and test local realism by violating the Mermin and Svetlichny inequalities.
These results represent a significant breakthrough for entangled multi-photon state production by eliminating the constraints of outcome post-selection, providing a novel resource for optical quantum information processing.
\end{abstract}

\maketitle

Quantum optical technologies promise to revolutionize fields as varied as computing, metrology and communication.
In most cases these applications require entangled states of light, but because photons are notoriously weakly interacting, creating entanglement between photons after they have been produced is challenging~\cite{Kok2007}.
Consequently, the ability to generate entanglement during the production process is of crucial importance for photons.
New capabilities of quantum sources are thus critical for the advancement of quantum optical implementations.

The production of high-quality multi-photon entanglement, such as Greenberger-Horne-Zeilinger (GHZ) states~\cite{Greenberger1990},  is particularly demanding.
Currently the most established method for producing photonic entanglement is spontaneous parametric downconversion (SPDC).
This is a process which naturally produces photons in pairs, making it simple to entangle the various degrees of freedom of two photons~\cite{Edamatsu2007}.
On the other hand, experiments with three or more entangled photons%~\cite{Bouwmeester1999,Pan2001,Eibl2004,Yao2012}
~\cite{Bouwmeester1999,Pan2001,Eibl2003,Eibl2004,Zhao2005,Walther2005,Lu2007,Yao2012,Pan2012}
have thus far relied on combining photons from two or more different pair sources using linear optics and employing outcome post-selection: selecting only a specific subset of measurement outcomes while ignoring others~\cite{Zeilinger1997}. With this approach, the action of observing the photons both creates and destroys the state at the same time.
While this post-selection may be acceptable for some applications, it restricts the usefulness of the resulting entangled states for others.
One example is heralding Bell states, also known as event-ready entanglement, which is the ability to know that a maximally entangled two-photon state is present before it is destroyed~\cite{Zukowski1993,Kok2000}.
This task, known to be useful for applications such as quantum repeaters~\cite{Briegel1998}, loophole-free Bell tests~\cite{Cabello2012} and optical quantum computing~\cite{Pittman2001,Browne2005}, is in theory easily achieved with an appropriate three-photon state but does not work if this state is created with SPDC and outcome post-selection.
Creating three-photon entanglement directly, without the need for such post-selection, would therefore represent a significant advance in photonic quantum information processing.

This goal can be achieved through cascaded downconversion~\cite{Greenberger1990,DAriano2000,Hubel2010}, a process where one of the photons from a primary SPDC process is used to pump a secondary downconversion source.
Specifically, if the primary source produces polarization-entangled photon pairs, and one of those photons is used to pump a secondary polarization-entangled source~\cite{Hubel2011}, the resulting three-photon state will be a GHZ entangled state (Fig.~\ref{Fig:Setup}a).
In this work, we use cascaded downconversion to produce entangled photon triplets directly without relying on outcome post-selection.
We fully characterize the entangled photon triplets with quantum state tomography, use them to perform local realism tests and to generate heralded Bell states.

Cascaded downconversion naturally produces photon triplets which are entangled in energy and time.
This was recently verified experimentally~\cite{Shalm2013}, under the assumption that downconversion conserves energy.
However,  the creation of more useful polarization-entangled photon triplets with cascaded downconversion, and verifying it conclusively, is significantly more challenging.
To produce the state, it is necessary to create a coherent superposition of two orthogonally polarized cascaded downconversion processes where the photons must be indistinguishable in their spectral, timing, and spatial characteristics.
In addition, the phase between the two processes has to be stable.
Meeting these requirements is especially demanding due to the properties of the high-efficiency downconverters employed here to make cascaded downconversion possible.

Fully characterizing a three-qubit state with quantum state tomography requires at least $27$ measurement settings. Performing this number of measurements with sufficient event statistics would be unfeasible with the highest previously reported detection rates of 7 triplets per hour~\cite{Shalm2013}, where an important limiting factor was the low single-photon detection efficiency. Here, we employ newly developed superconducting nanowire single-photon detectors (SNSPDs) with high system detection efficiency of over 90\% at 1550~nm~\cite{Marsili2013}, promising a hundred-fold increase in detected triplet rates. This dramatic improvement enables us to perform quantum state tomography and other demanding tests and applications of the three-photon entangled state.

\section*{State production and characterization}

\begin{figure}[htp]
\includegraphics[width=\linewidth]{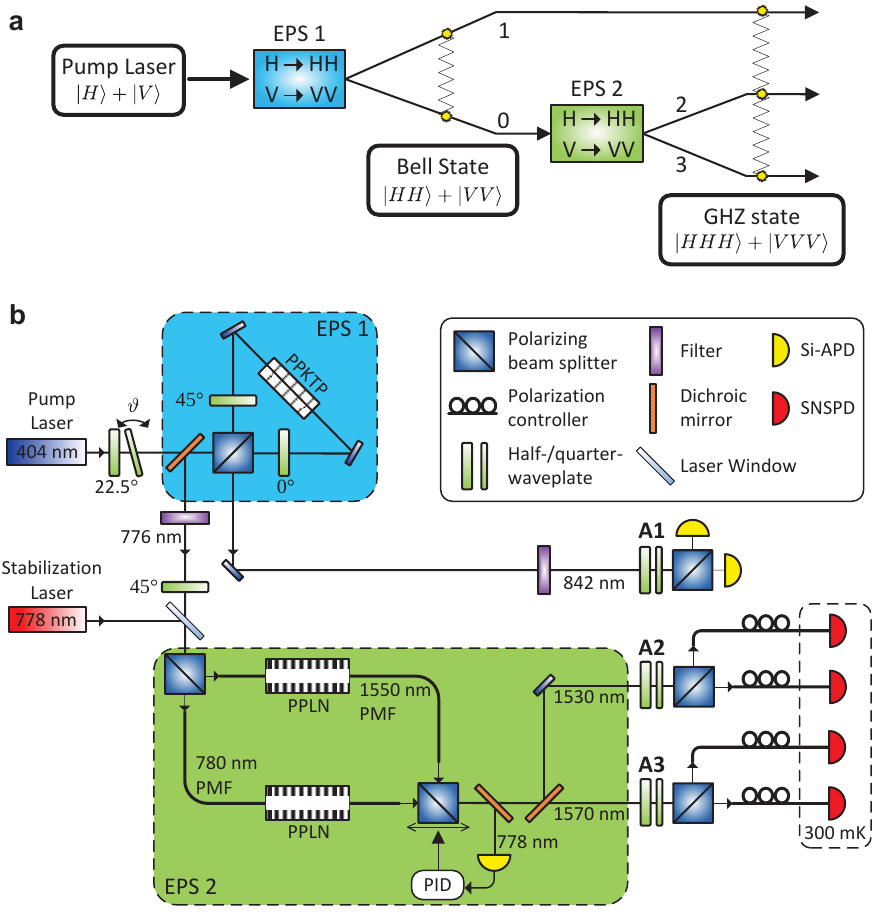}
\caption{\textbf{Polarization entangled photons using cascaded SPDC.} \textbf{a,} Schematic of the source. The first entangled photon source (EPS1) produces entangled photons in modes 0 and 1. The photon in mode 0 is used to pump the second entangled photon source (EPS2), thus transferring the entanglement to two new photons in modes 2 and 3 to produce a GHZ state.  \textbf{b,} Detailed setup of the experiment. A Sagnac source produces entangled photon pairs at 842~nm and 776~nm using a periodically-poled potassium titanyl phosphate (PPKTP) crystal.  The photons at 776~nm are used to pump a Mach-Zehnder source, which produces entangled photons at 1530~nm and 1570~nm in periodically-poled lithium niobate (PPLN) waveguides. The three-photon state is analyzed using controllable measurement settings implemented with motorized wave plates (A1, A2 and A3) and polarizing beam splitters. Photons at 842~nm are detected using silicon avalanche photodiodes (Si-APD), while photons at telecom wavelengths are detected using superconducting nanowire single-photon detectors (SNSPD). The signal from all detectors is sent to a time-tagging unit. The phase in the interferometer is controlled using a piezo-controller and a proportional-integral-derivative controller (PID). See Methods for additional details.}
\label{Fig:Setup}
\end{figure}

The three-photon states we aim to produce are GHZ states of the form
\begin{equation}
|\mathrm{GHZ}^\pm \rangle = \frac{1}{\sqrt{2}} (|HHH\rangle \pm |VVV\rangle ).
\label{GHZ_plus}
\end{equation}
Here $|H\rangle$ and $|V\rangle$ represent horizontally and vertically polarized photons respectively.
The setup (Fig.~\ref{Fig:Setup}b) can be understood as a cascade of two sources of entangled photon pairs~\cite{Hubel2011}.
First, a Sagnac source~\cite{Fedrizzi07,Kim2006} produces non-degenerate polarization-entangled photon pairs with wavelengths of 776~nm and 842~nm into modes 0 and 1 respectively.
These are in the Bell state $|\Phi\rangle=\frac{1}{\sqrt{2}}(|H\rangle_0|H\rangle_1+e^{i\theta(\vartheta)}|V\rangle_0|V\rangle_1)$, where the phase $\theta(\vartheta)$ can be controlled by tuning the tilt angle $\vartheta$ of the quarter-wave plate (QWP) in the pump beam.
The 776~nm photon is used to pump the second entangled photon pair source, a polarizing Mach-Zehnder interferometer with a downconversion crystal in each arm~\cite{Herbauts2013}.
If it downconverts, a pair of photons at 1530~nm and 1570~nm is created in modes 2 and 3 with a polarization state depending on the pump photon according to: $|H\rangle_0 \rightarrow |H\rangle_2|H\rangle_3$ or $|V\rangle_0 \rightarrow e^{i\phi} |V\rangle_2|V\rangle_3$, where $\phi$ is the phase difference between the paths.
This phase is kept constant using active stabilization.

\begin{figure}[htp]
\includegraphics[width=\linewidth]{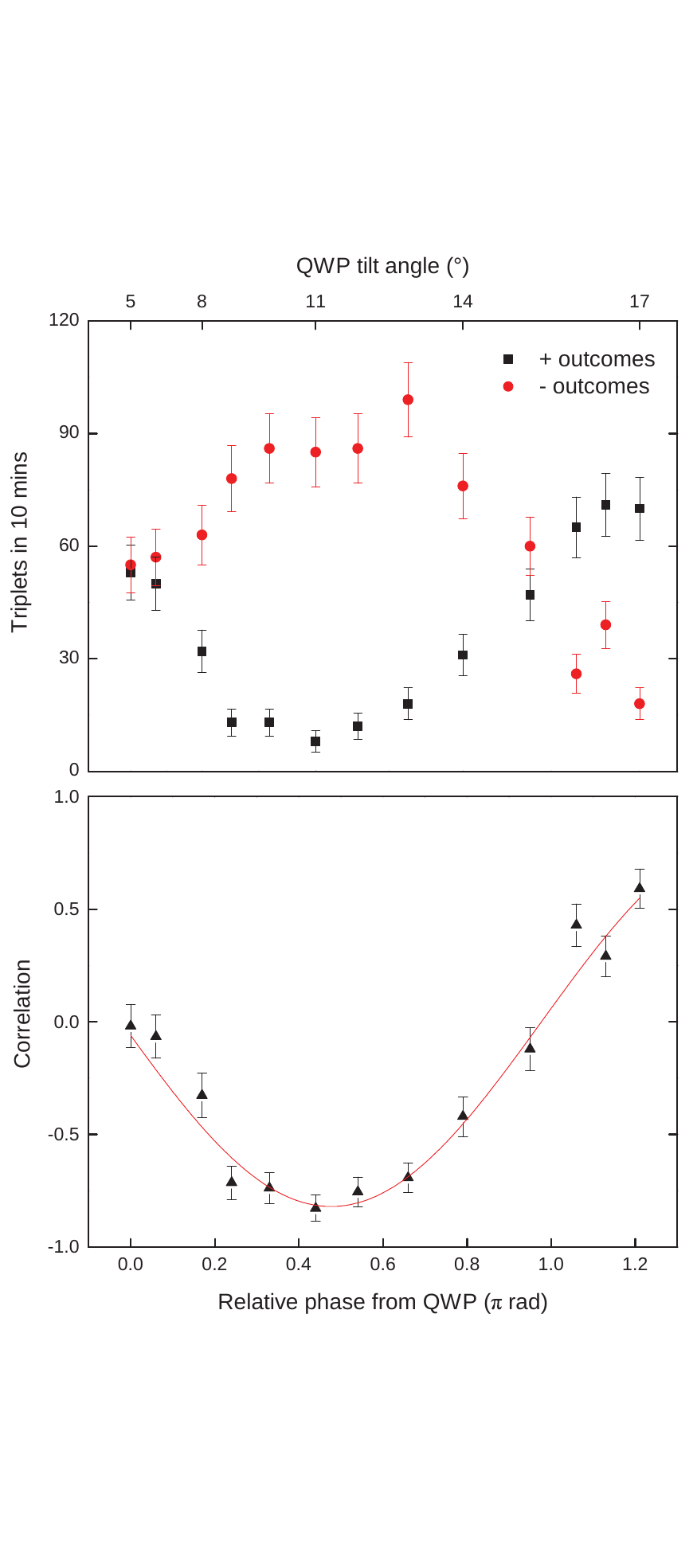}
\caption{\textbf{Measurement to determine optimal phase.} Measured triplets with positive (black squares) and negative (red circles) contributions to the diagonal basis correlation (top), and the corresponding correlation $E(\sigma_x , \sigma_x, \sigma_x)$ (bottom). The line is a sinusoidal fit with the amplitude and phase as fitting parameters, from which we extract an amplitude of $0.82 \pm 0.03$. Setting the QWP tilt angle $\vartheta$ to $11^{\circ}$ produces a relative phase of $(0.44\pm0.03)\pi$ and minimizes the correlation, resulting in a $|\mathrm{GHZ}^- \rangle $ state.}
\label{Fig:Scan_phase}
\end{figure}

The quantum state describing the photon triplets is
\begin{equation}
|\psi \rangle = \frac{1}{\sqrt{2}} \left(|H\rangle_1|H\rangle_2|H\rangle_3 + e^{i [\phi+\theta(\vartheta)]}|V\rangle_1|V\rangle_2|V\rangle_3\right).
\label{eq:GHZ}
\end{equation}
\noindent Each photon is subjected to a projective polarization measurement, and detected using single-photon detectors. All three-photon detection events are recorded as a set of time stamps. To analyze the data we use a coincidence window of 1.25~ns, which is larger than the combined timing jitters of the detectors and the timing electronics. We observe the phase dependance of the GHZ states by changing the pump QWP tilt angle $\vartheta$, and measuring the three-photon correlation in the diagonal polarization basis, $E(\sigma_x,\sigma_x,\sigma_x)$  with $\sigma_x=|H\rangle \langle V| + |V\rangle \langle H|$ (see Methods).
Quantum mechanics predicts that for the state in equation~\ref{eq:GHZ}, $E(\sigma_x,\sigma_x,\sigma_x)=\langle\sigma_x\sigma_x\sigma_x\rangle=\cos[\phi+\theta(\vartheta)]$.
The high-visibility sinusoidal dependance of the correlation on the phase $\theta(\vartheta)$ (Fig.~\ref{Fig:Scan_phase}) is a first signature of GHZ entanglement.
For subsequent measurements, the QWP tilt angle is set such that the correlation is either at a minimum or a maximum, resulting in $|\mathrm{GHZ}^- \rangle$ or $|\mathrm{GHZ}^+ \rangle $ respectively.

\begin{figure}[htp]
\includegraphics[width=\linewidth]{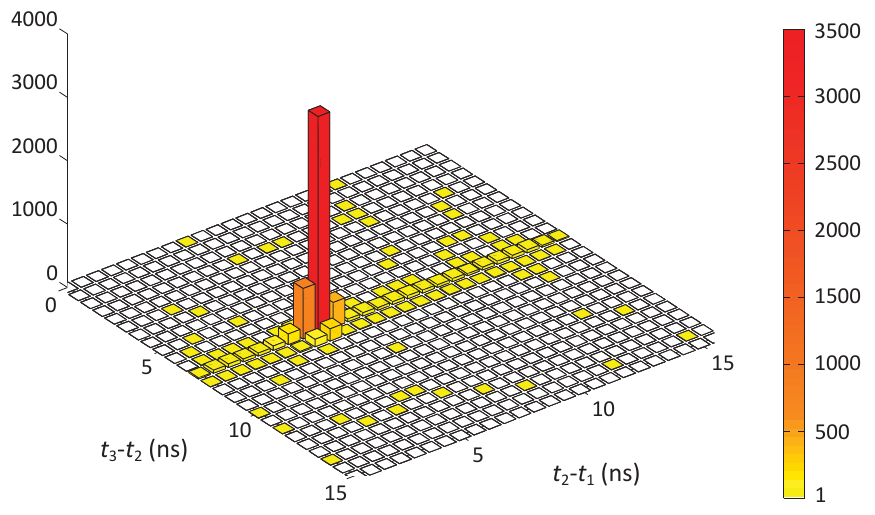}
\caption{\textbf{Two-dimensional histogram of time differences between detected photon events.} The large peak corresponds to photon triplets from cascaded down-conversion, showing that they have tight time-correlations.
The line above the background at $t_3-t_2\approx7~\mathrm{ns}$ is the main source of accidental triplets. It is due to events where a photon pair produced in the second downconversion is detected within 15~ns of an unrelated photon at 842~nm. The reason a similar line is not seen for a constant value of $t_2-t_1$ is that the count rates at detectors 2 and 3 are three orders of magnitude smaller than those at $D_1$, so an accidental three-fold coincidence is much more likely to involve an uncorrelated photon at $D_1$. The resulting signal to noise ratio in this histogram is 73:1.}  \label{Fig:Histogram}
\end{figure}

\begin{figure}[htp]
\includegraphics[width=3.4in]{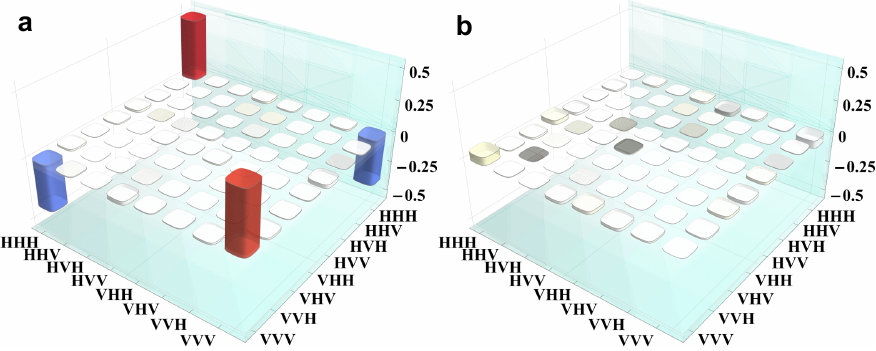}
\caption{\textbf{Real (a) and imaginary (b) parts of the reconstructed three-photon density matrix.} The density matrix is reconstructed from the measured three-fold coincidences with no background subtraction.}
\label{Fig:GHZ_tomo}
\end{figure}

We fully characterize the three-photon state by performing quantum state tomography to reconstruct its density matrix.
All 27 possible combinations of $\sigma_x$, $\sigma_y=-i|H\rangle \langle V| + i |V\rangle \langle H| $ and $\sigma_z=|H\rangle \langle H| - |V\rangle \langle V|$ are applied to the three photons.
Each setting is measured for 16 minutes, over a total of $7.2$ hours.
A histogram of time differences for all three-photon detections (Fig.~\ref{Fig:Histogram}) shows the the tight temporal correlations of the triplets.
Using the coincidence window of 1.25~ns around the peak yields a total of 4798 three-fold coincidences which are used for tomographic state reconstruction.
This corresponds to a detection rate of 11.1 triplets per minute.
The state is reconstructed using a semidefinite-programming algorithm implementation of the maximum likelihood method~\cite{James2001}.
The reconstructed density matrix $\rho$ (Fig.~\ref{Fig:GHZ_tomo}) has a fidelity with the state $|\mathrm{GHZ}^- \rangle$ of $F=\langle\mathrm{GHZ}^-|\rho | \mathrm{GHZ}^- \rangle = 86.2\%$, a fidelity with $|\mathrm{GHZ}^+ \rangle$ of $10.9\%$ and a purity $P=\mathrm{Tr}(\rho^2)$ of 0.776.
The most likely sources of imperfection are an incomplete overlap of the downconversion spectra of the two crystals in the second entangled pair source and an imperfect phase stabilization in the Mach-Zehnder interferometer.
Nonetheless, our GHZ state fidelity is, to the best of our knowledge, the highest fidelity for a three-photon GHZ state measured with tomography, surpassing the previous records of $84\%$~\cite{Lavoie2009,Lu2011b}.

\section*{Local realism tests}
The original motivation behind the introduction of GHZ states was that multipartite entanglement allows for a striking demonstration of the incompatibility of local realism and quantum mechanics~\cite{Greenberger1990}, through inequalities such as Mermin's~\cite{Mermin1990} or Svetlichny's~\cite{Svetlichny1987}.
These inequalities can also be recast as entanglement criteria~\cite{Toth2005b,Bancal2011,Cereceda2002}, and here we use them as a further demonstration of the quality of our GHZ state.

The Mermin inequality is derived by imposing locality and realism for all three particles.
Here, we look at the following three-particle version of the inequality.
\begin{small}
\begin{equation*}
S_\mathrm{Mermin} = |E(a',b',c) +  E(a',b,c') + E(a,b',c') - E(a,b,c)| \leq 2
\label{eq:Mermin}
\end{equation*}
\end{small}
The inequality holds for any local hidden variable theory.
It can be violated with a GHZ state by applying the measurements $a=b=c=\sigma_x$ and $a'=b'=c'=\sigma_y$, in the ideal case reaching the arithmetic limit of 4.
These measurements are a subset of those used for the three-photon tomography.
Of the 4798 triplet counts from the tomography, 674 correspond to the measurements for the Mermin inequality.
They lead to the correlation values shown in table~\ref{tab:MerminSvet}.
Combining these results in a Mermin parameter of $\langle S_\mathrm{Mermin} \rangle = 3.04 \pm 0.10$ that violates the local hidden variable limit by 10 standard deviations.
Because we use Pauli measurements in the Mermin inequality, its violation is also a confirmation that the state is genuinely tripartite entangled~\cite{Toth2005b}.
The same conclusion can be reached even if we do not implement ideal Pauli measurements, since a Mermin parameter larger than $2\sqrt{2}$ is a device-independent test of genuine tripartite entanglement~\cite{Bancal2011}.

\begin{table}[htp]
    \renewcommand{\arraystretch}{0.5} %make table single spaced in double-spaced environement
    \centering % used for centering table
     \begin{tabular}{l l | l l}
    \hline
    \multicolumn{2}{c}{Mermin} &  \multicolumn{2}{c}{Svetlichny}\\
    \hline
    $E(a,b,c)$ &  $ -0.78 \pm  0.05 $    &$E(a,b,c)$ &  $ ~~0.56 \pm  0.06 $ \\
    $E(a,b',c')$ &  $ ~~0.74 \pm  0.05 $    &$E(a,b,c')$ &  $ ~~0.63 \pm  0.06 $  \\
    $E(a',b,c')$ & $ ~~0.74 \pm  0.05  $    &$E(a,b',c)$ &  $ ~~0.65 \pm  0.06$ \\
    $E(a',b',c)$ &    ~~$ 0.77 \pm  0.05 $   &$E(a,b',c')$ &  $ -0.55 \pm  0.06$ \\
    &                                   &$E(a',b,c)$ &  $ ~~0.59 \pm  0.06$\\
    &                                   &$E(a',b,c')$ &  $ -0.59 \pm  0.06$\\
    &                                   &$E(a',b',c)$ &  $ -0.62 \pm  0.05$\\
    &                                   &$E(a',b',c')$ &  $ -0.71 \pm  0.05$\\
    \hline
    $S_\mathrm{Mermin}$ & $~~3.04 \pm 0.10$ &    $S_\mathrm{Svet}$ & $~~4.88 \pm 0.16$  \\
    \hline
    \end{tabular}
    \caption{Mermin and Svetlichny correlations. Note that the Mermin and Svetlichny measurements for $c$ and $c'$ are not the same.} %title of Table
    \label{tab:MerminSvet} % is used to refer this table in the text
 \end{table}

What the Mermin inequality cannot do is confirm the presence of tripartite nonlocality as it can be maximally violated with models allowing for arbitrarily strong correlations between two of the particles~\cite{Collins2002,Cereceda2002}.
The Svetlichny inequality addresses this problem, by allowing for arbitrarily strong correlations between any pair of particles, but otherwise enforcing locality and realism~\cite{Svetlichny1987}.
A violation of the Svetlichny inequality thus guarantees the presence of multipartite nonlocality~\cite{Bancal2013}, and rules out a large class of non-local hidden variable theories which Mermin's inequality cannot.

The Svetlichny inequality for three particles is
\begin{small}
\begin{equation*}
\begin{split}
S_\mathrm{Svet} = |E(a,b,c) +  E(a,b,c') + E(a,b',c) - E(a,b',c') \\
   +  E(a',b,c)  - E(a',b,c') - E(a',b',c) - E(a',b',c')| \leq 4.
\end{split}
\end{equation*}
\end{small}

\noindent This inequality can be violated with a GHZ state, but the Pauli measurements from the three-qubit tomography are no longer sufficient.
To test the Svetlichny inequality we perform another experiment using the measurement settings $a=b=\sigma_x$, $a'=b'=\sigma_y$, $c=\frac{1}{\sqrt{2}}(\sigma_x+\sigma_y)$ and $c'=\frac{1}{2}(\sigma_x-\sigma_y)$. In the ideal case, these measurements would result in a value of $S_\mathrm{Svet}=4\sqrt{2}$, which is the maximum value allowed by quantum mechanics.
For this experiment 1960 three-fold coincidences are measured over a period of 3.2 hours.
The values of the correlation are shown in Table~\ref{tab:MerminSvet}.
We find a Svetlichny parameter of $\langle S_\mathrm{Svet}\rangle=4.88 \pm 0.16$, violating the bound by 5 standard deviations.
To the best of our knowledge, this is the strongest measured violation of the three-particle Svetlichny inequality to date.

\section*{Heralded Bell states}
Our cascaded downconversion method of generating GHZ states allows us to directly herald Bell states.
This important task cannot be done with sources based on two independent SPDC sources and linear optics, including those which have been used to post-select GHZ correlations~\cite{Bouwmeester1999}.
Indeed, a minimum of six photons from SPDC are required~\cite{Kok2000,Sliwa2003,Barz2010,Wagenknecht2010} .
This is an example of the superiority of directly created quantum states compared to those produced only in post-selection.

\begin{figure}[ht]
\includegraphics[width=\linewidth]{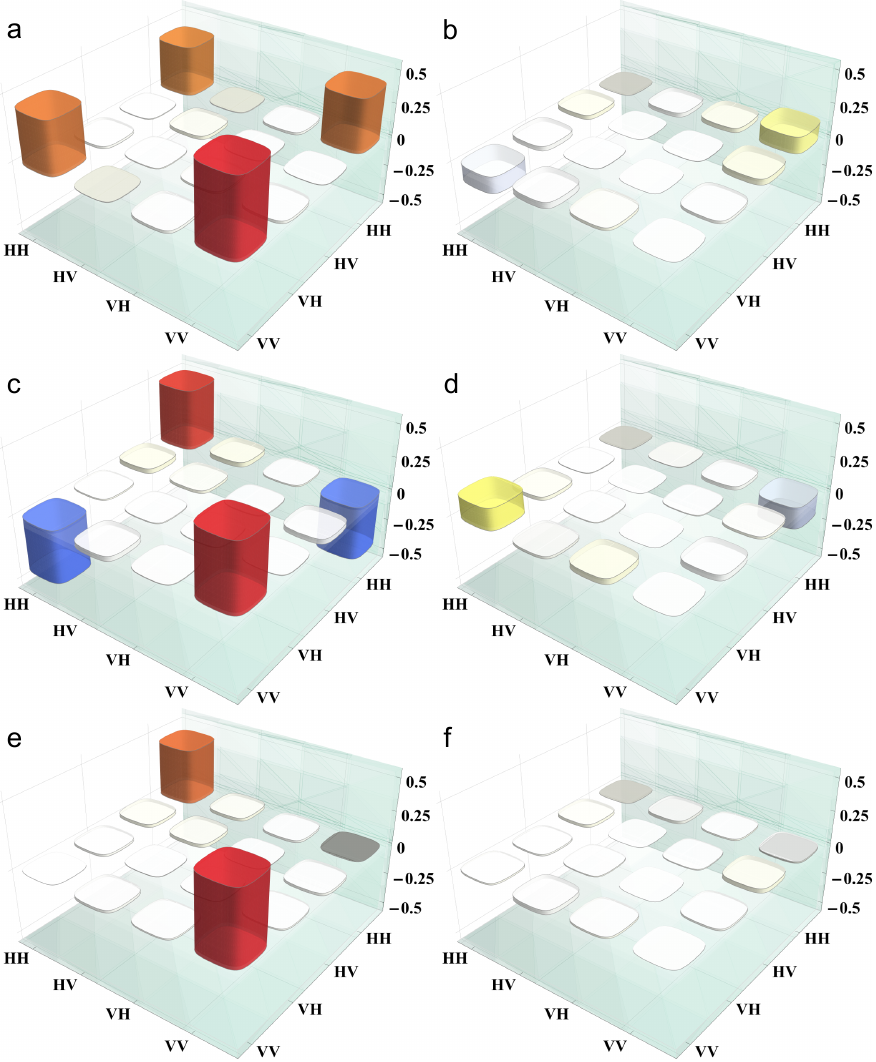}
\caption{\textbf{Real and imaginary parts of the reconstructed density matrices of the heralded two-photon states.} The density matrices are reconstructed from 1632 triplets which were measured in 3.6 hours. \textbf{a, b,} Heralding with $|D\rangle$ results in a state close to $|\Phi^+\rangle$. \textbf{c, d,} Heralding with $|A\rangle$ results in a state close to $|\Phi^-\rangle$.
\textbf{e, f,} When heralding with  $|D\rangle$ and $|A\rangle$ but ignoring the measurement outcomes, the coherent terms vanish, resulting in an incoherent mixture of $|HH\rangle$ and $|VV\rangle$.}
\label{Fig:HeraldedBellPairsTomo}
\end{figure}

To illustrate how our setup can be used as an event ready source of two-photon entanglement, we rewrite the GHZ state as
\begin{equation}
|\mathrm{GHZ}^+ \rangle = \frac{1}{\sqrt{2}} (|\Phi^+ \rangle |D\rangle + |\Phi^- \rangle|A\rangle),
\end{equation}
\noindent where $|D\rangle=\frac{1}{\sqrt{2}}(|H\rangle+|V\rangle)$ and $|A\rangle=\frac{1}{\sqrt{2}}(|H\rangle-|V\rangle)$ represent diagonal and anti-diagonal polarizations respectively, and $|\Phi^\pm \rangle= \frac{1}{\sqrt{2}} (|HH\rangle \pm |VV\rangle )$ are Bell states.
By projecting one of the photons in the diagonal basis, we can herald the presence of one of two Bell states in the other two modes.
The heralding detection should come from one of the photons at telecom wavelengths so that the conversion efficiency of the second downconversion does not affect the overall heralding efficiency.
In our experiment, we chose the 1530~nm photon to act as the herald.
For this measurement the phase is set to prepare a $|\mathrm{GHZ}^+ \rangle$ state.
We measure the 1530~nm photon in the diagonal basis, and perform quantum state tomography on the other two photons.
The density matrices resulting from each of the heralding outcomes (Fig.~\ref{Fig:HeraldedBellPairsTomo}, a to d) have a fidelity of 89.3\% with $|\Phi^+\rangle$ when heralding with $|D\rangle$, and 90.4\% with $|\Phi^-\rangle$ when heralding with $|A\rangle$.
Ignoring the outcome of the heralding measurement results in an incoherent mixture of $|HH\rangle$ and $|VV\rangle$ (Fig.~~\ref{Fig:HeraldedBellPairsTomo}, e and f).
The fidelity with an equally weighted incoherent mixture is 96.6\%.

We measure a rate of heralded two-photon pairs of 450 per hour.
From this we can extract the heralding efficiency of the system, defined as the probability of detecting a Bell state given a heralding signal.
The signal at the heralding detectors is dominated by dark counts, of which there are approximately 330 per second.
The resulting heralding efficiency is $1.9 \times 10^{-4}$.
However, this is not a fundamental limit; it is dominated by the ratio of signal photons to dark counts at the heralding detector.
In cases where the heralding signal is not caused by a dark count but by a signal photon, the heralding efficiency is as high as 0.06, limited by the coupling efficiency of the other two photons.

The measured heralding efficiency could be improved by using detectors with a lower dark count rate, by increasing the triplet production rate, or by optimizing the efficiency of single-mode fiber coupling.
Alternatively, switching the pump to a pulsed laser would provide additional timing information on the expected arrival time of the triplets, which could be exploited to remove a significant portion of the dark counts at the heralding detectors.

It is interesting to compare the performance of our source to previous experiments related to heralded Bell states.
Triggered entangled two-photons states have been produced with rates and fidelities similar to that of our setup from quantum dot systems~\cite{Akopian2006,Young2009}.
However, these experiments have much lower heralding efficiencies; to the best of our knowledge, the best reported heralding efficiency for these systems is $3.3 \times 10^{-9}$, five orders of magnitude lower than what we measure here~\cite{Salter2010}.

Experiments based on six-photon schemes resulted in two-photon states with a fidelity of 84\%~\cite{Barz2010}  and  87\%~\cite{Wagenknecht2010}.
The measured heralding efficiency of approximately $10^{-2}$ (including coupling and detection losses) reported by the six-photon experiments is higher, but with the changes discussed above our measured heralding efficiency would approach or even surpass this value.
In terms of detection rates, however, cascaded downconversion has a significant advantage; the six-photon experiments detected at most 4 heralded Bell states per hour, a rate that is two orders of magnitude less than what we measure with our cascaded SPDC source.
Moreover, the six-photon schemes have an inherent trade-off between trigger rates and heralding efficiency.
In our setup there is no such trade-off.
The heralding efficiency is entirely limited by experimental imperfections, and would in fact be improved by higher trigger rates.

Because of the high rate of heralded Bell states with cascaded SPDC, we are able to accumulate enough statistics to violate a Bell inequality with our heralded two-photon states.
We use the CHSH inequality~\cite{Clauser1969}
\begin{equation*}
S_\mathrm{CHSH} = |E(a,b) -  E(a,b') \pm E(a',b) \pm E(a',b') | \leq 2,
\label{eq:CHSH}
\end{equation*}
\noindent where $a=\sigma_z$, $a'=\sigma_x$, $b=\frac{1}{\sqrt{2}}(\sigma_z+\sigma_x)$ and $b'=\frac{1}{\sqrt{2}}(\sigma_z-\sigma_x)$.
Quantum mechanics predicts that the inequality can be violated up to a maximum of  $S_\mathrm{CHSH} = 2\sqrt{2}$.
The results of our measurements are shown in Table~\ref{tab:CHSH}.
We find $S_\mathrm{CHSH}=2.62 \pm 0.16$ when heralding with $|D\rangle$, and $S_\mathrm{CHSH}=2.70 \pm 0.19$ when heralding with $|A\rangle$.
Both correspond to violations by over 3 standard deviations of the local hidden variable limit.

 \begin{table}[htp]
    \renewcommand{\arraystretch}{0.5} %make table single spaced in double-spaced environement
    \centering % used for centering table
    \begin{tabular}{l l l }
    \hline 
    &  $|D\rangle$ Heralding &  $|A\rangle$ Heralding \\

    \hline

    $E(a,b)$ & $~~0.71 \pm 0.07$  & $~~0.77 \pm 0.06$   \\
    $E(a,b')$ & $-0.66 \pm  0.08$  & $-0.65 \pm 0.07$  \\
    $E(a',b)$ & $~~0.57 \pm 0.08$   & $-0.59  \pm 0.10$  \\
    $E(a',b')$ & $~~0.68 \pm 0.07$  & $-0.69  \pm  0.07$  \\ % [1ex] adds vertical space
    \hline 
    $S_\mathrm{CHSH}$ & $~~2.62 \pm 0.16$  & $~~2.70 \pm 0.19$  \\
    \hline
    \end{tabular}
    \caption{CHSH correlations and $S_\mathrm{CHSH}$ when heralding Bell states with $|D\rangle$ and $|A\rangle$.} %title of Table
    \label{tab:CHSH} % is used to refer this table in the text
 \end{table}

An important feature of our method of heralding two-photon entangled states is that the amount of entanglement of the resulting two-qubit state can be tuned based on the heralding measurement.
For example, by projecting the second photon of a $|\mathrm{GHZ}^- \rangle$ state onto $|\chi(\beta)\rangle=\cos{\beta}|H\rangle_2+\sin{\beta}|V\rangle_2$, we obtain states of the form
\begin{equation}
|\psi(\beta)\rangle = \cos{\beta}|H\rangle_1|H\rangle_3 - \sin{\beta}|V\rangle_1|V\rangle_3.
\end{equation}
To verify this, we vary the projection angle $\beta$, and perform tomography on the resulting two-photon state.
The fidelity of the measured states with the predicted states, for $\beta=\pi/4$, $\beta=\pi/8$ and $\beta=0$ is 78.4\%, 87.8\% and 96.4\% respectively (82.0\%, 85.6\% and 94.8\% for the orthogonal projections).
The density matrices for these states are shown in Fig.~\ref{Fig:NonMaxEntangled}.

\section*{Discussion}

In this experiment we demonstrate the direct generation of three-photon polarization entanglement with cascaded downconversion.
This method does not rely on the interference of independently produced photon pairs, or on outcome post-selection of detected photons; every photon triplet produced in our source is in the desired GHZ state.
The unique properties of this source enable a multitude of photonic quantum information tasks.
As a first such demonstration, we have shown that our source can herald high-fidelity Bell states.
It could also be useful as a source of multipartite entanglement for quantum communications protocols, such as quantum secret sharing~\cite{Hillery1999}.
We expect that our photon triplets are entangled in energy-time~\cite{Shalm2013}, opening the door to a demonstration of hyper-entangled photon triplets~\cite{Kwiat1997}.
With improved coupling efficiency out of the secondary downconversion, our method could be used for photon precertification~\cite{Cabello2012} to mitigate the impact of loss inherent to sending photons over long distances; this would allow for extended range of quantum communication, device-independent quantum key distribution~\cite{Gisin2010}, and loophole-free Bell tests~\cite{Cabello2012}.
In addition, with further improvements in conversion efficiency through novel materials or pumped third-order nonlinearities~\cite{Langford2011}, it may be possible to add more stages to the downconversion cascade.
This provides an avenue to the direct generation of entangled states of four or more photons, and consequently the heralding of GHZ states.

\begin{acknowledgments}
This work was financially supported by the Ontario Ministry of Research and Innovation ERA, QuantumWorks, NSERC, OCE, Industry Canada, CIFAR, CRC and CFI. We thank T. Zhao for contributions to the phase-stabilization software.
\end{acknowledgments}

\bibliography{Polarization_entangled_triplets_arXiv}

\section*{Methods}
\subsection*{Production of photon triplets}
A 25~mW 404~nm fiber-coupled grating-stabilized laser diode is used to pump a Sagnac source of entangled photons.
The downconversion occurs in a 30~mm periodically-poled potassium titanyl phosphate crystal.
The phasematching in the crystal is temperature tuned to produce entangled photon pairs at 776~nm and 842~nm.
The 842~nm photons are measured according to analyzer A1.
The 776~nm photons are sent into a polarizing Mach-Zehnder interferometer, which contains a 30~mm periodically-poled lithium niobate waveguide in each arm.
The PPLN waveguides are also temperature controlled, phasematched to produce photons centered at 1530~nm and 1570~nm.
After the Mach-Zehnder interferometer, the telecom photons are split by a dichroic mirror.
The photons at 1530~nm and 1570~nm are measured at analyzers A2 and A3 respectively.
The combined coupling and detection efficiency of the 842~nm photons is $\eta_1=0.23$, while for the 1530~nm and 1570~nm photons it is $\eta_2=\eta_3=0.30$, as measured from the ratio of photon detections to coincident photon detections.

\subsection*{Stabilization of the Mach-Zehnder interferometer}
The phase in the interferometer is kept constant by active stabilization.
A piezoelectric positioning stage %(Piezosystem Jena PX 50 CAP)
controls the position of the PBS at the exit of the Mach-Zehnder, based on a feedback signal provided by a 778~nm stabilization laser.
The piezoelectric statge's range alone is insufficient to stabilize the interferometer over long periods of time.
It is therefore mounted on a motorized linear stage %(Newport, MFA-CC),
which is activated whenever the piezoelectric stage approaches the limits of its range of motion.

\subsection*{Projective measurements}
The projective measurements on each photon are controlled using one half-wave plate and one quarter-wave plate, each of which is in a computer-controlled motorized rotation stage, placed in front of a polarizing beam splitter (PBS). Photons are detected at both outputs of the PBS. The correlation of a measurement is obtained by calculating the difference in relative frequency of events with a positive and negative product of outcomes.
For example, for the $\sigma_z \sigma_z \sigma_z$  measurement, the correlation value is explicitly given by:
\begin{tiny}
\begin{equation*}
E(\sigma_z  , \sigma_z,  \sigma_z  )  =  \frac{ N_{hhh}-N_{hhv}-N_{hvh}+N_{hvv}-N_{vhh}+N_{vhv}+N_{vvh}-N_{vvv}}{N_{hhh}+N_{hhv}+N_{hvh}+N_{hvv}+N_{vhh}+N_{vhv}+N_{vvh}+N_{vvv}}
\end{equation*}
\end{tiny}
where $N$ is the number of counts with each outcome combination, and $h$ and $v$ represent the positive and negative eigenvalue outcomes of the $\sigma_z$ measurement.

To mitigate the effect of any imbalance in coupling or detection efficiency, we set the wave plates to alternate which output represents the positive outcome of any given measurement. For example the $\sigma_z \sigma_z \sigma_z$ measurement is performed in eight different ways, where all three photons being transmitted by their respective PBSs corresponds to a projection onto the states:  $|HHH\rangle$, $|HHV\rangle$, $|HVH\rangle$, $|HVV\rangle$, $|VHH\rangle$, $|VHV\rangle$, $|VVH\rangle$ and $|VVV\rangle$.

\subsection*{Photon detection}
Two types of detectors are used for this experiment.
The 842~nm photons are detected with free-running silicon avalanche photodiodes (Si-APD) which have approximately $40\%$ efficiency at that wavelength.
The photons at 1530~nm and 1570~nm are detected using free-running tungsten silicide superconducting nanowire single-photon detectors (SNSPD) with  $90\%$ detection efficiency.
The SNSPDs are operated at a temperature of approximately 330 mK inside a compact, sealed, two-stage sorption-pumped ${}^3\mathrm{He}$ refrigerator with a sorption-pumped ${}^4\mathrm{He}$ stage for heat sinking of the wiring and optical fibers.
The sorption refrigerator stages are cooled by a closed-cycle Gifford-McMahon cryocooler with a nominal cooling power of 100~mW at 4.2~K.
For these experiments, the complete cycle lasts 6.5 hours, of which approximately 4 hours is spent with the detectors at operating temperature.
We record time stamps of all events when three photons are detected within 15~ns of each other using a time-tagger with a resolution of 156~ps.

\subsection*{Heralding non-maximally entangled states}
As was discussed in the main text, our two-photon state heralding method allows us to tune the amount of entanglement in the heralded states.
We start with the $|\mathrm{GHZ}^- \rangle$ state
\begin{equation*}
|\mathrm{GHZ}^- \rangle = \frac{1}{\sqrt{2}} (|HHH\rangle - |VVV\rangle ).
\end{equation*}
By projecting the heralding photon onto the state given by $|\chi(\beta)\rangle= \cos{\beta}|H\rangle + \sin{\beta}|V\rangle$, we obtain the following two-photon state in the other two modes:
\begin{equation*}
|\psi(\beta)\rangle = \cos{\beta}|HH\rangle - \sin{\beta}|VV\rangle.
\end{equation*}
\noindent To showcase this, we selected three values of $\beta$: $\pi/4$, $\pi/8$ and $0$. For each value of $\beta$, we reconstructed the heralded two-photon state using quantum state tomography. The reconstructed density matrices of the heralded states are shown in Fig.~\ref{Fig:NonMaxEntangled}. The fidelity with the ideal states for $\beta=\pi/4$, $\beta=\pi/8$ and $\beta=0$ is 78.4\%, 87.8\% and 96.4\% respectively for the positive outcome of the $|\chi(\beta)\rangle$ measurement, and 82.0\%, 85.6\% and 94.8\% for the negative outcome of the $|\chi(\beta)\rangle$ measurement.

\begin{widetext}

\begin{figure}[h]
\includegraphics[width=5.3in]{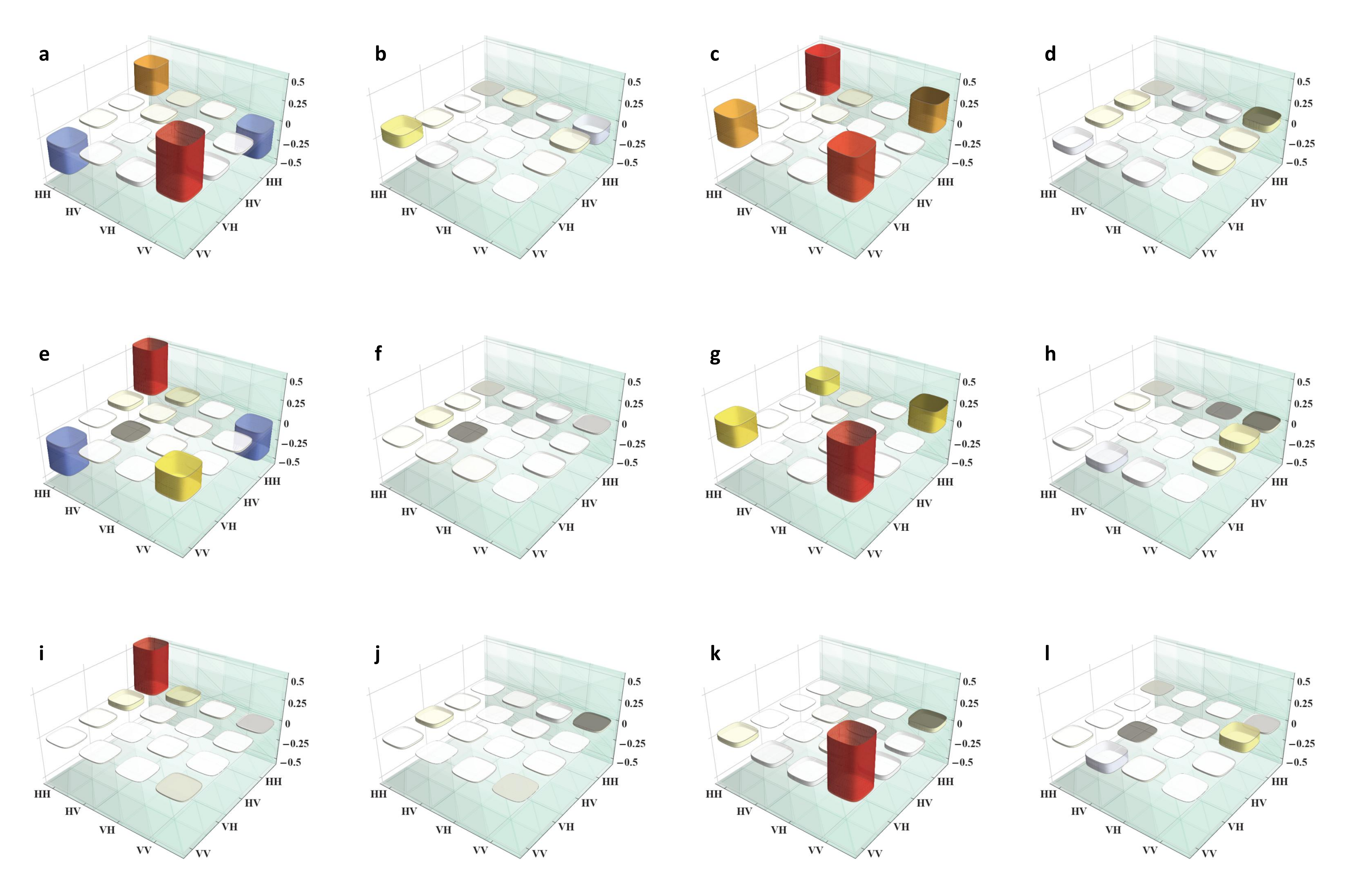}
\caption{\textbf{Reconstructed density matrices of heralded non-maximally entangled two-photon states.} The real and imaginary parts of the density matrices are shown when heralding with $\cos{\beta}|H\rangle + \sin{\beta}|V\rangle$ and the corresponding orthogonal measurement. For $\beta=\pi/4$, the fidelities with the ideal states are 78.4\% (\textbf{a} and \textbf{b}), and 82.0\% (\textbf{c} and \textbf{d}); for $\beta=\pi/8$ the fidelities are 87.8\% (\textbf{e} and \textbf{f}) and 85.6\% (\textbf{g} and \textbf{h}); and for $\beta=0$ the fidelities are 96.4\% (\textbf{i} and \textbf{j}) and 94.8\% (\textbf{k} and \textbf{l}).}
\label{Fig:NonMaxEntangled}
\end{figure}

\subsection*{Description of the quantum state}
Assuming a monochromatic pump of frequency $\omega_p$, the quantum state after the first downconversion can be written as
\begin{equation*}
\Psi_{SPDC1} \approx \int_{\omega_1}  d\omega_1 G_1(\omega_1,\omega_p-\omega_1) \left[ a_{1,H}^\dagger(\omega_1) a_{0,H}^\dagger(\omega_p-\omega_1) + e^{i \theta(\vartheta)}
a_{1,V}^\dagger(\omega_1) a_{0,V}^\dagger(\omega_p-\omega_1) \right] ,
\end{equation*}
\noindent where $G_1(\omega_1,\omega_p-\omega_1)$ is the joint-spectral functions resulting from the phasematching in the PPKTP crystal. We assume that it the same for the horizontal and vertical photons since in the Sagnac configuration they both come from the same crystal.
After the second SPDC, the state becomes
\begin{equation*}
\begin{split}
\Psi_{CSPDC} &\approx \int_{\omega_1} \int_{\omega_2} d\omega_1 d\omega_2 G_1(\omega_1,\omega_p-\omega_1)
 \bigg[  G_{2,H}(\omega_2, \omega_p-\omega_1 - \omega_2) a_{1,H}^\dagger(\omega_1) a_{2,H}^\dagger(\omega_2) a_{3,H}^\dagger(\omega_p-\omega_1 - \omega_2) \\ & + e^{i [\theta(\vartheta)+\phi]}
  G_{2,V}(\omega_2, \omega_p-\omega_1 - \omega_2) a_{1,V}^\dagger(\omega_1) a_{2,V}^\dagger(\omega_2) a_{3,V}^\dagger(\omega_p-\omega_1 - \omega_2) \bigg],
\end{split}
\end{equation*}
\noindent where $G_{2,H}(\omega_2, \omega_p-\omega_1 - \omega_2)$ and $G_{2,V}(\omega_2, \omega_p-\omega_1 - \omega_2)$ are joint-spectral functions for photons produced in either of the PPLN crystals. These include effects of phasematching, as well as any dispersion coming from the fibers after the downconversion.  If the two joint spectral functions are equal, in other words if the photons produced in either one of the two PPLN crystals are indistinguishable, then we have
\begin{equation*}
\begin{split}
\Psi_{CSPDC} &\approx \int_{\omega_1} \int_{\omega_2} d\omega_1 d\omega_2 G_1(\omega_1,\omega_p-\omega_1) G_{2}(\omega_2, \omega_p-\omega_1 - \omega_2) \\
  \bigg[  a_{1,H}^\dagger(\omega_1)& a_{2,H}^\dagger(\omega_2) a_{3,H}^\dagger(\omega_p-\omega_1 - \omega_2) +  e^{i [\theta(\vartheta)+\phi]} a_{1,V}^\dagger(\omega_1) a_{2,V}^\dagger(\omega_2) a_{3,V}^\dagger(\omega_p-\omega_1 - \omega_2)\bigg].
\end{split}
\end{equation*}
\end{widetext}
In this form, the energy-time and polarization correlations of the state are evident.
Considering only polarization, the resulting state is
\begin{equation*}
\begin{split}
\Psi_{CSPDC} = \frac{1}{\sqrt{2}} (|HHH\rangle + e^{i [\theta(\vartheta)+ \phi]} |VVV\rangle ),
\end{split}
\end{equation*}
\noindent which is the desired entangled GHZ state.

It is crucial that the two joint-spectral functions $G_{2,H}(\omega_2, \omega_p-\omega_1 - \omega_2)$ and $G_{2,V}(\omega_2, \omega_p-\omega_1 - \omega_2)$  be equal.
To achieve this, the phasematching curves of both crystals are measured, and their temperatures are independently controlled and set to have maximum overlap of the two downconversion spectra.
Additionally, because the telecom photons produced in the second downconversion are broadband ($\sim$30~nm FWHM), the entanglement visibility can be easily degraded by unbalanced group velocity dispersion in optical fibers.
This was significant in our experiment because the fibers pigtailed to the PPLN samples were not exactly of the same length.
\begin{figure}[!b]
\includegraphics[width=\linewidth]{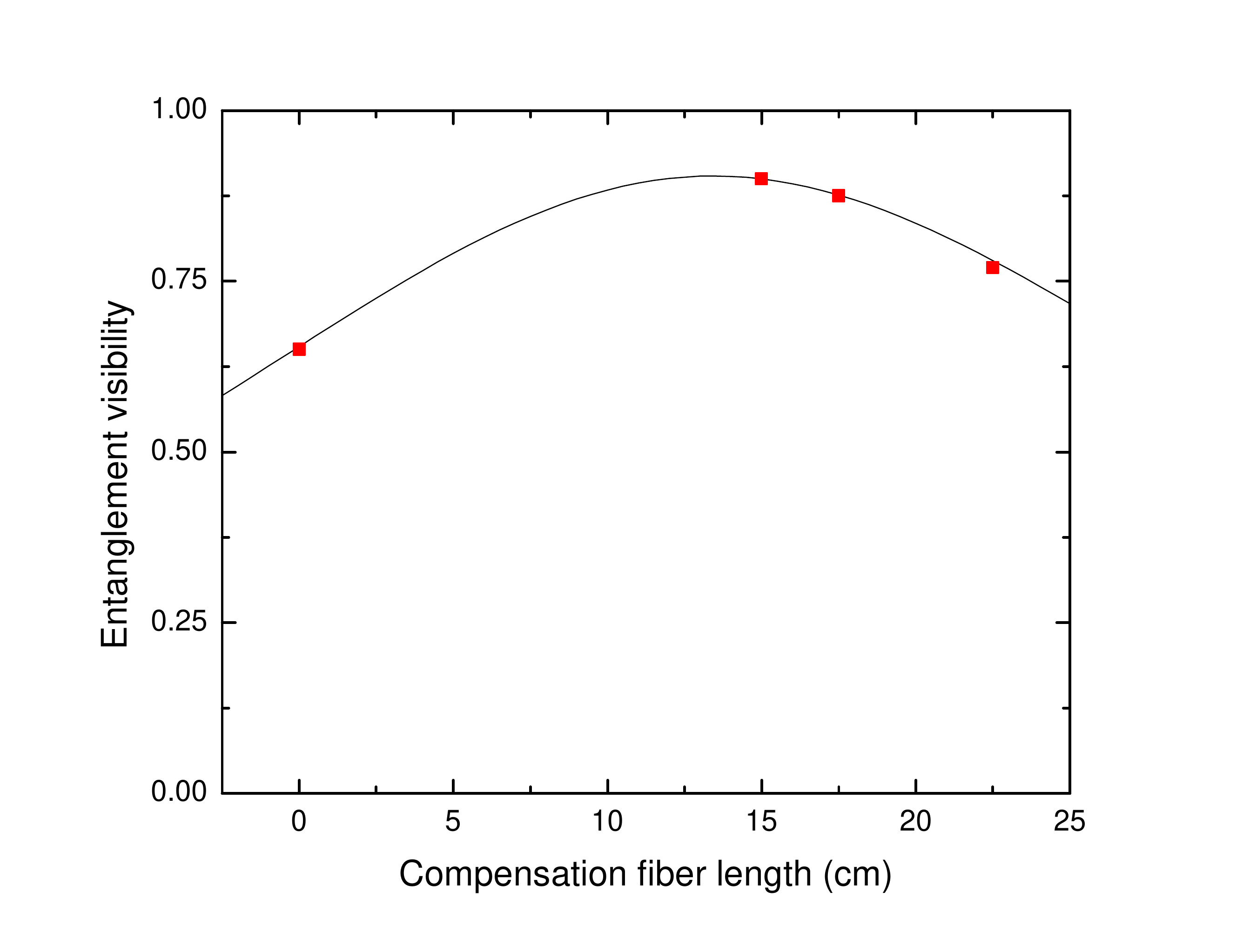}
\caption{\textbf{Effect of fiber length mismatch on entanglement visibility.} The data points correspond to measured entanglement visibility of the Mach-Zehnder source when different compensation fibers were inserted.
The line is calculated using dispersion data for fused silica, assuming an SPDC bandwidth of 28nm and a maximum visibility of $90\%$}
\label{Fig:dispersion}
\end{figure}
We compensated for this imbalance by including additional polarization maintaining fiber in one arm.
To find the optimal length of fiber to add, we tried different lengths of fibers and measured the resulting entanglement visibility of the Mach-Zehnder source, as shown in Fig.~\ref{Fig:dispersion}. This is measured by pumping the Mach-Zehnder source with a laser and measuring the photon pairs in the diagonal basis without stabilizing the phase.
The visibility of the resulting fringes is the entanglement visibility.
Based on these measurements, we added 15~cm of polarization maintaining fiber to one path of the interferometer.
\end{document}